\documentclass[12pt,a4paper]{article}
\usepackage{amsmath}
\usepackage{amssymb}

 \newcommand{\NN}{\mathbb N}
 \newcommand{\RR}{\mathbb R}
 
 \newcommand{\ZZ}{\mathbb Z}

 \newcommand{\QQ}{\mathbb Q}

 \newtheorem{theorem}{Theorem}
 
 \newtheorem{prop}{Proposition}
 
 \newtheorem{fact}{Fact}

 \input{epsf}
 
\begin{document}
 \bibliographystyle{unsrt}

 \parindent0pt

\begin{center}
{\large \bf A note on palindromicity} 

\vspace{8mm}
        
                 {\sc Michael Baake}\footnote{Heisenberg-Fellow}
 
\vspace{5mm}

   Institut f\"ur Theoretische Physik, Universit\"at T\"ubingen, \\

   Auf der Morgenstelle 14, D-72076 T\"ubingen, Germany  

\end{center}

\vspace{10mm}
\begin{abstract} 
Two results on palindromicity of bi-infinite words in a finite alphabet are
presented. The first is a simple, but efficient criterion to exclude
palindromicity of minimal sequences and applies, in particular, to the 
Rudin-Shapiro sequence.
The second provides a constructive method to build palindromic minimal
sequences based upon regular, generic model sets with centro-symmetric window.
These give rise to diagonal tight-binding models in one dimension
with purely singular continuous spectrum.

\end{abstract}

\parindent15pt
\vspace{15mm}

\subsection*{Introduction}

Since the discovery of quasicrystals some 15 years ago, there has been a renewed
interest in the spectral properties of non-periodic Schr\"odinger operators, and
in discrete versions of them in particular, see \cite{Suto} for a review. 
The best-studied case is the so-called diagonal tight-binding model, written as
an operator $L$ acting on the Hilbert space $\ell^2(\ZZ)$ as follows
\begin{equation}
      ( L^{}_x u)_n \; = \; u^{}_{n+1} + u^{}_{n-1} + x^{}_n u^{}_n
\end{equation}
where $x$ denotes a bi-infinite sequence of potentials taking finitely many
(pairwise different) real values. One key task is now to infer spectral properties 
of the operator $L^{}_x$ from properties of the sequence $x$.

Of specific interest is the case where $x$ is linked to the fixed point of
a primitive substitution rule, and a lot is known here, see 
\cite{Suto,Anton,HKS,Allouche}
and references therein for relevant contributions. One remarkable feature of
examples such as the Fibonacci or the Thue-Morse sequence is that the spectrum
of the corresponding operator is purely singular continuous, i.e.\ there is
neither a point spectrum nor an absolutely continuous spectrum present, or, in
more common terms, the generalized eigenstates are all critical, and neither
localized nor extended.

Clearly, this asks for generalizations and for simple criteria to decide upon
this property in other examples. This is precisely the starting point of a number
of important contributions, and also the starting point of the paper by Hof, Knill
and Simon \cite{HKS} where palindromicity of $x$ was singled out as one efficient
tool to decide upon the spectral nature, building on earlier work
Jitomirskaya and Simon \cite{JS}, see also \cite[Appendix]{HKS}.
Recall that a word is a palindrome if it reads backwards the same as
forwards, such as ``level'' or ``deed'' in ordinary language, and that
$x$ is called palindromic if it contains palindromes of arbitrary length.
With criteria built upon these concepts, the spectral nature of quite a number 
of popular examples can be understood.

However, one nagging example is, and always has been, the operator $L$ with a 
potential according to the Rudin-Shapiro sequence. Based upon numerical evidence, 
Hof et al.\ conjectured that the Rudin-Shapiro sequence
does not contain palindromes of arbitrary length, and hence that this sequence 
escapes another attempt to be spectrally classified. This conjecture was
soon after answered to the affirmative by Allouche \cite{Allouche} who linked
the Rudin-Shapiro sequence (and some generalizations of it) to paperfolding
sequences. As a result, he could show that the Rudin-Shapiro sequence can only
contain palindromes of length 1,2,3,4,5,6,7,8,10,12,14, see the remark at the
end of his paper.

It is the first aim of this note to provide an alternative proof of this statement.
The justification is mainly that the method is quite different, constructive,
and can easily be used for general primitive substitution rules, so it will
be helpful in other cases. In a certain sense, the approach shown below is
entirely straight-forward and follows easily from standard results in \cite{Queffelec},
but it seems to be largely unknown, and hence worthwhile being resurfaced.
Since this part is mainly combinatorial, we will recollect some notions
of ergodic theory only afterwards.

The main aim then is to use palindromicity to extend the known
class of diagonal tight-binding models with purely singular continuous spectrum
considerably. This is possible by adapting some standard results from the theory
of model sets to construct a rather general class of sequences $x$ which lead to
strictly ergodic (i.e.\ minimal and uniquely ergodic) systems $X$ and thus allow 
the application of the following
\begin{prop} \label{prop-hks}
 If $X$ is aperiodic, strictly ergodic and palindromic, there is a generic 
 $Y\subset X$ such that for $x\in Y$ the operator $L_x$ has purely singular 
 continuous spectrum.
\end{prop}
This is essentially \cite[Corollary 7.3]{HKS} except for the requirement of
aperiodicity which is missing in \cite{HKS}, though implicitly assumed
and, in fact, necessary.

\subsection*{Palindromes}

Let ${\cal A} = \{a^{}_1, a^{}_2, \ldots , a^{}_r\}$ be a finite set called 
{\em alphabet} and ${\cal A}^*$ the set of finite words in elements of $\cal A$.
A {\em palindrome} is a word that reads backwards the same as forwards.
The empty word is considered a palindrome. 
An element $x$ in ${\cal A}^{\ZZ}$ (bi-infinite sequence) or in 
${\cal A}^{\NN}$ (semi-infinite sequence) is called {\em palindromic} if it 
contains palindromes of arbitrary length. If a word $w$ occurs in a sequence $x$
as $w = x^{}_n x^{}_{n+1} \ldots x^{}_m$, we say that $w$ is centred 
or positioned at $(n+m)/2$.

An element $x\in {\cal A}^{\ZZ}$ is called {\em strongly palindromic}, with parameter 
$B$, if there exists a sequence $w_i$ of palindromes of length $\ell_i$ centred at
$m_i$ such that $|m_i| \rightarrow \infty$ together with 
$e^{B |m_i|}/\ell_i \rightarrow 0$ for $i \rightarrow \infty$. This means that
there is a subsequence of palindromes of diverging length {\em and} position
such that the lengths still grow faster than an exponential of the positions.
This somewhat strange property plays a key role in a sufficient criterion for
purely singular continuous spectra. If $x$ is strongly palindromic, it is
clearly palindromic, but the converse need not be true.

Now, the later treatment of the Rudin-Shapiro sequence rests upon the 
following, rather trivial observation
\begin{fact} \label{prop1}
  Let $x$ be an infinite sequence of symbols. 
  If there exists an integer $n\geq 1$ such that
  no palindromes of length $n$ and none of length $n+1$ exist, there are no
  palindromes of length $m$ for any $m \geq n$, and $x$ is not palindromic.
\end{fact}
{\sc Proof}: Let $w$ be a finite string of symbols, of length $m$.
If $w$ is a palindrome, we can chop off one symbol at its beginning and
its end, and obtain another palindrome, this time of length $m-2$.
By assumption, the sequence $x$ does not contain any substring $w$ of length $n$
or $n+1$ which is a palindrome. If it contained one of length $m > n+1$, we 
iteratively chop off 2 symbols, one at the beginning and one at the end, 
until we obtain a palindrome either of length $n$ or of length $n+1$:
a contradiction. \hfill $\square$

So, this opens a route to establish non-palindromicity by brute force, provided
we find an easy and exhaustive way to catalogue all occuring substrings of $x$
of length $n$, until we eventually find the assumption of the previous Proposition
satisfied. This is actually sufficiently easy for substitution sequences, and
this is where we deviate from Allouche's approach \cite{Allouche}.

\subsection*{Substitution sequences}

Consider $\cal A$ as above, with $|{\cal A}| = r$.
An $r$-letter substitution rule is now a mapping $\sigma$ which attaches to
each element of $\cal A$ a word of finite length in the letters of $\cal A$,
\begin{equation}
         a^{}_j \; \mapsto \;  w^{}_j = \sigma(a^{}_j)
\end{equation}
where we assume that the word $w^{}_j$ has length $\ell_j$, i.e.\ consists
of $\ell_j$ consecutive letters. 
For several purposes, one needs the corresponding substitution matrix $M_{\sigma}$. 
Its entries are \cite{Queffelec}
\begin{equation}
        (M_{\sigma})_{ij} \; = \; \mbox{number of occurrences of 
             $a^{}_i$ in the word $w^{}_j$} \, .
\end{equation}
This convention leads to  $M_{\sigma\circ\varrho} = M_{\sigma} M_{\varrho}$.  
Sometimes the transpose is used, see \cite{BGJ} for details and a formulation
in the setting of free groups and their homomorphisms, followed by an
abelianization.

There are various interesting subclasses of substitutions. We are interested in
{\em primitive} ones which are characterized by the property that there is an
integer $k$ such that $M^k$ has (strictly) positive entries only.
We are interested in (bi-infinite) fixed points of a primitive substitution
rule $\sigma$. After possibly replacing $\sigma$ by some power of it, such
a fixed point always exists \cite{Queffelec}.

Up to now, things are pretty standard and appear in many texts. What is
less well known is the fact that such substitutions also induce substitutions
on words of length $N$, and one standardized way to do this properly is the
following. Let 
$\sigma(w)=a^{}_{i^{}_1} a^{}_{i^{}_2} a^{}_{i^{}_3} \ldots a^{}_{i^{}_n}$
be the substitution image of an arbitrary word $W$ of length $N$. Define
\begin{equation}  \label{induced}
   \sigma^{}_{N} (w) \; = \; 
             (a^{}_{i^{}_1} a^{}_{i^{}_2} \ldots a^{}_{i^{}_N})
             (a^{}_{i^{}_2} a^{}_{i^{}_3} \ldots a^{}_{i^{}_{N+1}})
             \ldots
             (a^{}_{i^{}_m} a^{}_{i^{}_{m+1}} \ldots a^{}_{i^{}_{m+N-1}})
\end{equation}
where $m$ is the length of $\sigma(a)$ if $a$ is the first letter of $w$.
Consequently, $m+N-1\leq n$ so that $\sigma^{}_{N}$ is well defined. Note that
this definition is designed to avoid double counting of strings of length $N$, 
see \cite{Queffelec} for details.

The relevant question is how to obtain a complete list ${\cal A}^{}_N$ of
all words of length $N$ that occur in the (by assumption existing !) fixed point 
$u$ of the primitive substitution, $\sigma(u)=u$. Since $\sigma$ is primitive, 
$\sigma^{}_{N}$ is also primitive when viewed as a substitution on the new ``alphabet''
${\cal A}^{}_N$, see \cite[Lemma V.12]{Queffelec}. There are now basically two
straight-forward methods to determine ${\cal A}^{}_N$, one based on the repetitivity of
$u$, and one on the nature of $\sigma^{}_{N}$. Let us start with a description of the 
latter.

The possible words of length $N$ certainly form a subset of those words that are 
obtained from possible words of length $N-1$ by adding to them any of the letters of 
our alphabet ${\cal A}={\cal A}^{}_1$. Let us denote this superset of potential words 
by ${\cal B}^{}_{N}$. The induced substitution $\sigma^{}_{N}$ can now be used to
determine ${\cal A}^{}_{N}$ from ${\cal B}^{}_{N}$.
If $w$ is a word of length $N$, let $\tilde{\sigma}^{}_N(w)$ denote the {\em set}
of words of length $N$ in $\sigma^{}_N(w)$ as they appear on the right hand
side of Eq.~(\ref{induced}). Clearly, 
$\tilde{\sigma}^{}_N({\cal B}^{}_{N})\subset{\cal B}^{}_{N}$, and after finitely many
iterations of $\tilde{\sigma}^{}_N$ we arrive at a stable set $C$, so that
$\tilde{\sigma}^{}_N(C)=C$. This means that the substitution $\sigma^{}_N$ 
itself is irreducible on $C$, hence primitive by the previous remarks, 
and $C={\cal A}^{}_{N}$. Usually, one actually finds 
$C=\tilde{\sigma}^{}_N({\cal B}^{}_{N})$, 
so only one iteration is required in this process.

Starting with ${\cal A}^{}_1$, we may use this procedure to determine ${\cal A}^{}_N$
iteratively up to any (finite) $N$ we wish -- a simple recursive program will do.
In fact, for typical examples, it is no problem to explicitly do it up
to $N=100$ say, and this is often sufficient to find the conditions to apply
Fact \ref{prop1} and rule out palindromicity. 

The other method mentioned to determine ${\cal A}^{}_N$ relies on a special property
of fixed points of primitive substitutions. If $\sigma(u)=u$, 
any word of length $\ell$ occurs in every substring of $u$ of length $L=L(\ell)$, and
$L$ grows only linearly in $\ell$. So, $L(\ell)\leq c \ell$ and one can estimate $c$
from the explicit form of $\sigma$. Then, to obtain ${\cal A}^{}_N$, one cuts out of
$u$ an arbitrary subword of length $cN$ and collects the different substrings of length
$N$. This is usually even quicker than the previous method, but requires the
determination of $c$.

But whatever method one prefers, the exclusion of palindromicity is usually very
efficient. Let us demonstrate this with an example.

\subsection*{The Rudin-Shapiro sequence}

Let $a(n)$ be the number of (possibly overlapping) blocks of type $11$ in the
binary expansion of $n$, for $n\geq 0$. If we define
\begin{equation}
     x^{}_n \; := \; {1\over 2} (1 - (-1)^{a(n)} ) \, ,
\end{equation}
we obtain the Rudin-Shapiro sequence as the half-infinite sequence
$(x^{}_0, x^{}_1, ... )$, i.e.\ as
\begin{equation} \label{rs-seq}
     0 \; 0 \; 0 \; 1 \; 0 \; 0 \; 1 \; 0 \; 0 \; 0 \; 0 \; 1 \; 1 \; 1\; \ldots
\end{equation}
One well-known alternative way is through a primitive\footnote{The 3rd power of
the substitution matrix has all entries positive.} 4-letter substitution rule,
\begin{equation}
     \sigma \, : \;
      \left( \begin{array}{ccc}
              a & \mapsto & ab \\
              b & \mapsto & ac \\
              c & \mapsto & db \\
              d & \mapsto & dc       \end{array} \right) \, ,
\end{equation}
which gives the sequence $a\,b\,a\,c\,a\,b\,d\,b\,a\,b\,a\,c\,d\,c ...$, followed 
by the mapping $\varphi$ which sends $a$ and $b$ to $0$ and $c$ and $d$ to $1$. 
This brings us back to (\ref{rs-seq}). The 4-letter version may be called
{\em quaternary} or 4-letter RS-sequence, while (\ref{rs-seq}) is the {\em binary} 
or ``classic'' RS-sequence.

So, we can use the methods from substitution rules, and determine the atlas of
substrings of length $n$. If we subject this atlas to the mapping $\varphi$, we
obtain the atlas of the Rudin-Shapiro sequence and can then, eventually, apply
Fact \ref{prop1}. We summarize the result in Table \ref{tab1}.

\begin{center}
\begin{table}
$$\begin{array}{|r|rr|rr|} 
\hline 
n  & \#^{(4)}(n) & \;{\rm P ?} & \#^{(2)}(n) & \;{\rm P ?} \\
\hline 
 1 &   4 &  {\rm yes}\;    &   2 &  {\rm yes}\; \\    
 2 &   8 &   {\rm no}\;    &   4 &  {\rm yes}\; \\
 3 &  16 &  {\rm yes}\;    &   8 &  {\rm yes}\; \\
 4 &  24 &   {\rm no}\;    &  16 &  {\rm yes}\; \\
 5 &  32 &  {\rm yes}\;    &  24 &  {\rm yes}\; \\
 6 &  40 &   {\rm no}\;    &  36 &  {\rm yes}\; \\
 7 &  48 &  {\rm yes}\;    &  46 &  {\rm yes}\; \\
 8 &  56 & {\rm \bf no}\;  &  56 &  {\rm yes}\; \\
 9 &  64 & {\rm \bf no}\;  &  64 &   {\rm no}\; \\
10 &  72 &                 &  72 &  {\rm yes}\; \\
11 &  80 &                 &  80 &   {\rm no}\; \\
12 &  88 &                 &  88 &  {\rm yes}\; \\
13 &  96 &                 &  96 &   {\rm no}\; \\
14 & 104 &                 & 104 &  {\rm yes}\; \\
15 & 112 &                 & 112 & {\rm \bf no}\;  \\
16 & 120 &                 & 120 & {\rm \bf no}\;  \\
17 & 128 &                 & 128 &   \\
18 & 136 &                 & 136 &   \\
19 & 144 &                 & 144 &   \\
20 & 152 &                 & 152 &   \\
\hline 
\end{array}$$
\caption{Number of words of length $n$ (complexity) in quaternary and binary 
Rudin-Shapiro sequence for $1\leq n\leq 20$, and status of palindromicity.}
\label{tab1}
\end{table}   
\end{center}

If $n\geq 8$, the complexity of both versions is the same. It is known \cite{AS}
to be $\#(n)=8n-8$. The 4-letter version has only palindromes of length 1,3,5
and 7, while the reduction to the binary version creates some new palindromes,
whence we get some of length 1,2,3,4,5,6,7,8,10,12 and 14, but not beyond.
Consequently, neither version is palindromic.

\subsection*{Further preliminaries and recollections}

{}For the sequel, we need some notions from symbolic dynamics and ergodic 
theory, see \cite{Queffelec} and \cite{Petersen} for details. Two elements 
$x,y$ of ${\cal A}^{\ZZ}$ are {\em locally indistinguishable} (LI) if
each substring of $x$ also occurs in $y$ and vice versa. This is an equivalence
relation and LI($x$) denotes the class represented by $x$, called LI-class.
In view of this, we call an LI-class palindromic if one (and hence any) 
representative is palindromic. Also, LI($x$) is called {\em aperiodic} if
$x$ is aperiodic. Consequently, an aperiodic LI-class does not contain
any periodic member.

A sequence $x$ is called {\em repetitive} if each substring of $x$ re-occurs in
$x$ in a relatively dense way, i.e.\ the distance between any two consecutive 
occurences is bounded\footnote{Note that some authors call this property ``minimal'' 
(which will show up in a different meaning shortly) or ``almost-periodic'' (which has
even more other meanings), wherefore I prefer ``repetitive'' in this context.}.
Examples of repetitive LI-classes are obtained by fixed 
points of primitive substitution rules (see above), but also by generic model sets 
(as we shall see later). 

The (two-sided) {\em shift} $S$ on ${\cal A}^{\ZZ}$ is defined by 
$(S(x))^{}_n := x^{}_{n+1}$. The {\em orbit} of $x\in {\cal A}^{\ZZ}$ under the 
shift is ${\cal O}(x) = \{ S^n x \mid n\in\ZZ\}$. Its closure in the product
topology is called the {\em orbit closure} of $x$ and denoted by
$\overline{{\cal O}(x)}$. A compact shift-invariant subset $X$ of
${\cal A}^{\ZZ}$ is called {\em minimal} if $\overline{{\cal O}(x)} = X$ 
for all $x\in X$. The following result is standard and follows directly from 
Gottschalk's Theorem \cite[Thm.\ 4.1.2]{Petersen}:
\begin{prop} \label{gs}
  $\;$ If $x$ is in ${\cal A}^{\ZZ}$, the following statements are equivalent:
  \begin{itemize}
    \item[(1)] \quad $x$ is repetitive.
    \item[(2)] \quad $\overline{{\cal O}(x)}$ is minimal.
    \item[(3)] \quad $\overline{{\cal O}(x)} = {\rm LI}(x)$.
  \end{itemize}
\end{prop}

With this result, we can rephrase Prop.\ 2.1 of \cite{HKS} as follows.
\begin{prop} \label{uncount}
  $\;$ Let $x\in {\cal A}^{\ZZ}$ be repetitive and palindromic.
  If $x$ is periodic, all elements of ${\rm LI}(x)$ are strongly palindromic.
  If $x$ is aperiodic, ${\rm LI}(x)$ contains uncountably many strongly palindromic 
  sequences.
\end{prop}
Strong palindromicity of periodic sequences is trivial.
The proof of the existence of uncountably many strongly palindromic sequences
in \cite{HKS} is constructive, but the statement is weaker than it appears,
because this could still be a thin set in LI($x$). For some recent work on extensions
we refer to \cite{Dama1,Dama2}.

\subsection*{1D model sets and derived sequences}

To keep things simple, we restrict to the following cut and project scheme
for a model set in one dimension. By definition,
this consists of a collection of spaces and mappings: 

\begin{equation}\label{cutandproject}
  \begin{array}{ccccc}
   \RR & \stackrel{\pi^{}_1}{\longleftarrow} & \RR \times \RR^n &
           \stackrel{\pi^{}_2}{\longrightarrow} & \RR^n  \\
    & & \cup & & \\ & & \tilde{L} & & \end{array}
\end{equation}
where $\RR$ and $\RR^n$ are two real spaces, $\pi^{}_1$ and 
$\pi^{}_2$ are the canonical projection maps onto them, and 
$\tilde{L} \subset \RR \times \RR^n$ is a lattice. 
We assume that $\pi^{}_1|^{}_{\tilde{L}}$ is injective and
that $\pi^{}_2(\tilde{L})$ is dense in $\RR^n$. We call $\RR$
(resp.\ $\RR^n$) the physical (resp.\ internal) space. We will assume that
$\RR$ and $\RR^n$ are equipped with Euclidean metrics and that
$\RR \times \RR^n$ is the orthogonal sum of the two spaces. 

A cut and project scheme involves, then, the projection of
a lattice into a space of smaller dimension, but a lattice that is 
transversally located with respect to the projection maps involved.

Let $L := \pi^{}_1(\tilde{L})$ and let
\begin{equation}\label{star}
     (\,)^* \, : \quad L \; \longrightarrow \; \RR^n
\end{equation}
be the mapping $\pi^{}_2 \circ (\pi^{}_1|^{}_{\tilde{L}})^{-1}$.
This mapping extends naturally to a mapping on the rational span 
$\QQ L$ of $L$, also denoted by $(\,)^*$.  The lattice $\tilde{L}$ can 
now also be written as
\begin{equation}
    \tilde{L} \; = \; \{ (z,z^*) \mid z \in L \}.            
\end{equation}
Now, let $\Omega \subset \RR^n$. Define
\begin{equation}
     \Lambda \; = \; \Lambda(\Omega) \; := \; 
                     \{ z \in L \mid z^* \in \Omega \, \} \, .
\end{equation}
We call such a set $\Lambda$ a {\em model set} (or {\em cut and project set}) 
if the following two conditions are fulfilled,
\begin{itemize}
\item[\bf W1] $\Omega \subset \RR^n$ is compact.
\item[\bf W2] $\Omega \;=\; \overline{\mbox{int}(\Omega)} \;\neq\;\emptyset$.
\end{itemize}
In addition, $\Lambda$ is called {\em regular}, if $\Omega$ is Riemann measurable, 
i.e.\ if
\begin{itemize}
\item[\bf W3] The boundary of $\Omega$ has Lebesgue measure 0,
\end{itemize}
and it is called non-singular or {\em generic}, if
\begin{itemize}
\item[\bf W4] $L^* \cap \partial\Omega = \emptyset$.
\end{itemize}

Let us just mention that there is no need to restrict to Euclidean internal
spaces, and one can, with little extra complication, extend the setup to
locally compact Abelian groups instead, which widens the class of structures 
covered considerably \cite{Moody,BMS,Martin1}. In any case, the extension
beyond the Euclidean codimension one situation is very natural, both physically
and mathematically. This is certainly needed for the model set description of
more general Pisot substitution (e.g.\ on alphabets of more than two letters), 
i.e.\ substitution rules whose inflation multiplier is a Pisot-Vijayaraghavan
number \cite{BMS}. Also, it quite naturally provides a huge class of other
model sets that cannot be described by a local substitution, but gives rise to
interesting tight binding models, as we shall see.

The key aspect of interest here is that generic, regular model sets are aperiodic 
and repetitive \cite{Martin1}, and any finite patch that occurs in the set does so 
with a positive uniform frequency \cite{Hof,Martin1}. Also, the model set is a 
Delone set of finite type,
and this means in particular that only finitely many different distances
between two consecutive points exist, $r$ say. If we attach $r$ different
letters to these intervals, we have mapped the model set to a bi-infinite
sequence $x$ in $r$ letters. Any finite word in it then occurs with a positive,
uniform frequency, and LI($x$) is strictly ergodic w.r.t.\ the action of $\ZZ$.
So we have
\begin{prop} \label{map}
   Let $\Lambda$ be a generic, regular model set in one dimension.
   Then, mapping different intervals between consecutive points of
   $\Lambda$ to different letters gives rise to a bi-infinite sequence 
   $x=x^{}_{\Lambda}$ in a finite alphabet that is strictly ergodic
   w.r.t.\ the action of $\ZZ$.
\end{prop}

Let $x\in{\cal A}^{\ZZ}$. We say that LI($x$) has {\em generalized inversion 
symmetry} if also $Rx$ is in LI($x$), where $(Rx)_i := x_{-i}$, and {\em strict
inversion symmetry} if there is some $y\in$LI($x$) so that $Ry=S^m y$ for
some $m\in\ZZ$, where $(Sx)_i = x_{i+1}$ is the shift. Clearly, if LI($x$) is
strictly inversion symmetric, then there must be a $y$ with either $Ry=y$ or
$Ry=Sy$, and LI($x$) is palindromic. The converse is also true if $x$ is repetitive.
\begin{theorem} \label{thm1}
   Let $x\in{\cal A}^{\ZZ}$ be repetitive. Then $x$, and hence ${\rm LI}(x)$, is 
   palindromic if and only if $\,{\rm LI}(x)$ is strictly inversion symmetric.
\end{theorem}
{\sc Proof}: In view of the previous remark, we only have to show that palindromicity
of $x$ implies strict inversion symmetry of LI($x$). If $x$ is palindromic, it
contains palindromes $w_i$ of length $\ell_i$, centred at $m_i$, with
$\ell_{i+1} > \ell_i$. Define $n_i=m_i$ if $m_i$ integer and $n_i=m_i+\frac{1}{2}$
if not. Now, consider the sequence $S^{n_i}x$ where the $i$th
element has a palindrome of length $\ell_i$ centred at 0 or at $1/2$. 
Compactness of ${\cal A}^{\ZZ}$ guarantees that there is a subsequence that 
converges to an infinite 
palindrome, i.e.\ to a $y$ with either $Ry=y$ or $Ry=Sy$. Since $x$ is repetitive,
LI($x$)$=\overline{{\cal O}(x)}$ by Prop.~(\ref{gs}), and $y$ lies in LI($x$).
\hfill $\square$

Now, if we start with a generic regular model $\Lambda$ set and attach the 
corresponding letter sequence $x^{}_{\Lambda}$ to it, we obtain a palindromic 
LI-class if the original model set was strictly inversion symmetric (which is 
defined by the requirement that $-\Lambda = \Lambda + t$ for some $t\in\RR$). 
This can easily be achieved by choosing
a window $\Omega$ that is inversion symmetric and has the property [{\bf W4}]
that $\partial\Omega \cap L^* = \emptyset$. For a given cut and project setup
with inversion symmetric window, this is clearly the generic case.
But we can still extend the situation in two ways.

First of all, the window clearly need not be inversion symmetric with respect to
the (completely artificial) origin of internal space, it is sufficient if it is
centro-symmetric, i.e.\ if $\Omega = -\Omega + c$ for some $c\in\RR^n$.
Also, the condition [{\bf W4}] can be replaced by a weaker one. Assume that
[{\bf W4}] is violated. Then, some shifted version $\Omega+c$ will observe it
again and produce a generic, regular model set. Now consider the corresponding
LI-class defined by it. It follows from the so-called {\em torus parametrization} 
\cite[App.\ A.1]{torus} that this LI-class contains at least $2^{n+1}$ inversion 
symmetric members, with equality holding if and only if all of them are
generic. So, even if the original condition [{\bf W4}] fails for $\Omega$, we
can still constructively check the other possibilities which correspond to specific 
shifts of the window. 

This situation is actually met in the standard example of
the Fibonacci chain: here, 4 fixed points on the torus exist, three of which
correspond to generic members and the fourth to a pair of singular members, see
\cite{torus} for details. To summarize, we apply Propositions \ref{map},
\ref{uncount} and \ref{prop-hks} and Theorem \ref{thm1} to obtain
\begin{theorem}
   Let $\Lambda$ be a regular, generic model set that is strictly inversion 
   symmetric, and let $x=x^{}_{\Lambda}$ be the corresponding aperiodic bi-infinite
   letter sequence. Then, LI($x$) contains an uncountable
   (and even generic) subset $Y$ with the
   property that the tight-binding operators $L_y$ on $\ell^2(\ZZ)$ in the sense of 
   Eq.~(1) has purely singular continuous spectrum for all $y\in Y$.
\end{theorem}

\subsection*{Concluding remarks}

There are many repetitive, palindromic LI-classes to which the
original argument by Hof, Knill and Simon can be applied, and examples obtained
by substitution rules or by the codimension one projection method with the
``standard" window constitute only a thin set in comparison. So, the 
appearance of purely singular continuous spectra in the 1D diagonal tight
binding model is also more common than originally anticipated. 

An open question, however, still is to what extent this spectral type is shared 
by other members of the same LI-class of potentials, see \cite{Dama4} for a
survey. As can be seen from recent
results on the complexity of palindromes in substitution generated sequences
\cite{Dama3}, it is unlikely that the method of strong palindromicity is able to 
settle this in general, and other ideas seem to be needed here.

\subsubsection*{Acknowledgements}

It is my pleasure to thank Jean-Paul Allouche, David Damanik, Robert V.\ Moody 
and Martin Schlott\-mann for helpful and clarifying discussions.
This work was supported by the German Science Foundation (DFG).

\end{document}